\documentclass[twocolumn,pre]{revtex4}
\usepackage{epsfig}
\usepackage{dcolumn}
\usepackage{bm}
 
\def\be{\begin{equation}}
\def\ee{\end{equation}}

\begin{document} 

\title{Slow dynamics in gelation phenomena: From chemical gels to colloidal 
glasses}
\author{Emanuela Del Gado,$^{a,d}$, Annalisa Fierro,$^{b,d}$, 
Lucilla de Arcangelis, $^{c,d}$ and Antonio Coniglio $^{b,d}$}
\affiliation{
${}^a$Laboratoire des Verres, Universit\'e Montpellier II, 34095 Montpellier, 
France}  
\affiliation{${}^b$Dipartimento di Scienze Fisiche, Universit\`a di Napoli
"Federico II",\\ Complesso Universitario di Monte Sant'Angelo,
via Cintia 80126 Napoli, Italy}
\affiliation{${}^c$Dipartimento di Ingegneria dell'Informazione, 
 Seconda Universit\`a di Napoli, via Roma 29, 81031 Aversa (Caserta), Italy}
\affiliation{${}^d$ INFM Udr di Napoli and Gruppo coordinato SUN}
\date{\today}
\begin{abstract}
We here discuss the results of $3d$ MonteCarlo simulations of a 
minimal lattice model for gelling systems. 
We focus on the dynamics investigated by means of the time autocorrelation 
function of the density fluctuations and the particle mean square displacement.
We start from the case of chemical gelation, i.e. with permanent bonds, and 
characterize the critical dynamics as determined by the formation of the 
percolating cluster, as actually observed  
in polymer gels. By opportunely introducing a finite bond lifetime 
$\tau_b$, the dynamics displays relevant changes and eventually the onset of a 
glassy regime. This has been interpreted in terms of a crossover to dynamics 
more typical of colloidal systems and a novel connection between classical 
gelation 
and recent results on colloidal systems is suggested. By systematically 
comparing the results in the case of permanent bonds to finite 
bond lifetime, the crossover and the glassy regime can be understood 
in terms of effective clusters. 
\end{abstract}
\maketitle

\section{Introduction}
The gelation transition transforms a viscous liquid into an elastic disordered 
solid. In general this is due to the formation in the liquid phase 
of a spanning structure, which makes the system able to bear stresses.\\ 
In polymer systems, this is due to chemical bonding, that can be induced in 
different ways \cite{flory,degl}, producing a polymerizations process. 
As firstly recognized by Flory, the change in the viscoelastic 
properties is directly related to the constitution inside the sol of a 
macroscopic polymeric structure, that characterizes the gel phase. 
In experiments \cite{viscoela} the viscosity coefficient grows as a power law  
as function of the relative difference from the critical polymer 
concentration with a critical exponent $k$. 
The onset of the elastic response in the system, as function of the same 
control parameter, displays a power law increasing of the elastic modulus 
with a critical exponent $f$. 
As implicitly suggested in the work of Flory and Stockmayer \cite{flory}, 
the percolation model is considered as the basic model for the chemical 
gelation transition and the macromolecular stress-bearing structure in these
systems is a percolating network \cite{degl,stau,adconst}.  
In experiments the gelling solution typically displays slow dynamics: 
the relaxation functions present a long time stretched exponential decay 
$\sim e^{-(\frac{t}{\tau_0})^{\beta}}$ as the gelation threshold is approached.
In particular at the gel point the relaxation process becomes critically slow, 
and the onset of a power law decay is  observed \cite{relax}.

In many other physical systems where aggregation processes and structure 
formation take place,  gelation phenomena can be observed. Typically, these are 
colloidal systems, i.e. suspensions of mesoscopic particles interacting 
via short range attraction. These systems are intensively investigated due to 
their relevance in many research fields 
(from proteins studies to food industry). 
Due to the possibility in experiments of opportunely 
tuning the features of the interactions, they also play the role of 
model systems.

In these systems strong attraction gives rise to a diffusion 
limited cluster-cluster aggregation process and may produce gel formation 
(colloidal gelation) at very low density as a spanning structure is formed 
\cite{dinsweitz}. 
The latter is generally quite different from the polymer gels case 
\cite{dlca} however, for what the viscoelastic behavior is concerned, 
this gelation transition closely resembles the chemical one, observed in 
polymer systems \cite{weitzna}. With a weaker attraction at higher densities a 
gelation characterized by a glass-like kinetic arrest \cite{relaxco,mallamace} 
may be observed. The relaxation patterns closely recall the 
ones observed in glassy systems and are well fitted 
by the mode-coupling theory predictions for super-cooled 
liquids approaching the glass transition \cite{goe}. On the theoretical 
side the application of the mode-coupling theory to systems with short range 
attractive interactions \cite{mct,ema,cates} ({\em attractive glasses}) 
has been recently considered and the connection with the colloidal glass 
transition has been proposed. The short range attraction enhances the caging 
mechanism characteristic of glassy regimes in hard sphere systems and produces
a glassy behavior at lower densities, depending on the attraction strength.

At lower densities, the role of the structure formation, as directly 
observed in some systems \cite{weitzclu}, might be relevant in the 
dynamics but it has not been clarified yet. 
Also the eventual crossover to the glassy regime \cite{weitzna}, 
as the density is increased, is not completely understood.
In this paper we investigate the 
connection among colloidal gelation, colloidal glass transition and 
chemical gelation. Some preliminary studies have been reported in 
\cite{dedecon5}.

We have considered a minimal model for gelling systems and performed extensive 
numerical simulations on $3d$ cubic lattices. In Sect. \ref{one} we give the 
details of the model and the simulations. In Sect.s \ref{two} and 
\ref{three} the results relative to relaxation and diffusion properties are 
presented and discussed. In the last section some
concluding remarks are given.

\section{Description of the model}
\label{one}
\subsection{Permanent bonds}
\label{oneperm}
Our model consists in a solution of monomers diffusing on a cubic lattice. 
As in most experimental  polymer gels,
we choose the monomers to be tetrafunctional.
Each monomer occupies a lattice elementary cell, and therefore eight vertices on
the lattice. To take into account the excluded volume interaction, two monomers 
cannot occupy nearest neighbors and next nearest neighbors cells on the 
lattice, i.e. nearest neighbor monomers cannot have common sites.
At $t=0$ we fix the fraction $\phi$ of present monomers with respect to the 
maximum number allowed on the lattice, and randomly quench bonds between them. 
This actually corresponds to the typical chemical gelation process that can be 
obtained by irradiating the monomeric solution. 
We form at most
four bonds per monomer, randomly selected along lattice directions and
between monomers that are nearest neighbors and next nearest neighbors 
(namely bond lengths $l=2,~ 3$).  Once formed, the bonds are permanent. 

For each value of $\phi$ there is an average value $N_{b}(\phi)$ of the 
fraction of formed bonds with respect to all the possible ones, 
obtained by averaging over different initial configurations.

Varying $\phi$ the system presents a percolation transition at
$\phi_{c} = 0.718 \pm 0.005$ \cite{dedecon3}. 
The critical exponents found at the transition agree with the random 
percolation predictions \cite{staul} (e.g. for the mean cluster size 
$\gamma \simeq 1.8 \pm 0.05$ and for the correlation length 
$\nu \simeq 0.89 \pm 0.01$ in $3d$ \cite{dedecon3}).

The monomers diffuse on the lattice via random local movements and the bond 
length may vary but not be larger than $l_{0}$ according to bond-fluctuation 
dynamics (BFD) \cite{carkr}, where the value of $l_{0}$ is determined by the 
self-avoiding walk condition. On the cubic lattice we have $l_{0}=\sqrt{10}$ 
in lattice spacing units and the allowed bond lengths are 
$l=2,\sqrt{5},\sqrt{6},3,\sqrt{10}$ \cite{carkr2}. 
We let the monomers diffuse to reach the stationary state and then study the 
system for different values of the monomer concentration.  

This lattice model with permanent
bonds has been introduced to study the critical behavior of the viscoelastic properties at the 
gelation transition\cite{dedecon}. 
The relaxation time is found to diverge 
at the percolation threshold $\phi_{c}$ with a power law behavior 
\cite{dedecon3}. 
The elastic response in the gel phase has been studied by means of the 
fluctuations in the free energy and goes to zero at $\phi_{c}$ with a power 
law behavior as well \cite{dedecon4}. 

\subsection{Bonds with finite lifetime}  
Colloidal gelation is due to a short range attraction and in general 
the particles are not permanently bonded. To take into account this crucial 
feature we introduce a novel ingredient in the previous model by considering 
a finite bond lifetime $\tau_{b}$ and study the effect on the dynamics. 

The features of this model with finite
$\tau_{b}$ can be realized in a microscopic model: 
a solution of monomers
interacting via an attraction of strength $-E$ and excluded volume repulsion.
Due to monomer diffusion the aggregation process eventually takes place.
The finite bond lifetime $\tau_{b}$ is related to the  attractive interaction
of strength $-E$, as $\tau_{b} \sim e^{E/KT}$.

We start with the same
configurations of the previous case, with a fixed $\phi$ where the bonds have
been randomly quenched as described above. During the monomer diffusion with
$BFD$ at every time step we attempt to break each bond with a frequency
$1/\tau_{b}$. Between monomers separated by a distance less than $l_{0}$ bonds
are then formed with a frequency $f_{b}$ \cite{nota3}. 
In order to obtain monomers 
configurations that are similar to 
the ones with permanent bonds, for each value of 
$\tau_{b}$ we fix $f_{b}$ so that the fraction of present bonds 
coincides with its average value in the case of permanent bonds, 
$N_{b}(\phi)$ \cite{nota1}.

With respect to the case of permanent bonds we notice that, 
as the finite bond lifetime $\tau_{b}$ corresponds to an attractive 
interaction of range $l_{0}$, it actually introduces a correlation in the bond 
formation and may eventually lead to a phase separation between a low density 
and a high density phase: 
There is no evidence of phase separation for the values of $\tau_{b}$ 
and $f_{b}$ considered in this paper. 
This is evident in Fig.\ref{fig1}, 
where typical equilibrium configurations with $\phi=0.6$ are shown in two 
different cases: in Fig.\ref{fig1}$(a)$ we have the case considered in this 
paper,  
obtained with $\tau_{b}=100$ and $f_{b}=0.02$; 
in Fig.\ref{fig1}$(b)$
with $\tau_{b}=2$ and $f_{b}=1$ the phase 
separation may occur.
The choice of monomers of functionality $4$, also in this case of finite bond 
lifetime, 
may correspond to a directional effect in the interaction \cite{nota2}.

The case of extremely large $\tau_{b}$ gives rise to different situations 
depending on the initial condition and the bond creation
process.
We consider two extreme cases:
I) Start with the initial configuration where the
monomers are randomly distributed, and the bonds are randomly 
quenched. At later times the frequency of forming bonds is zero.
This case corresponds to the permanent bond case (chemical gelation), 
described in section \ref{oneperm}. II) Start with a random
configuration of monomers and let the monomers diffuse and form bonds with 
infinite lifetime and frequency $f_{b}$. 
This phenomenon of irreversible aggregation, with the occurrence of gelation
after a spanning cluster is formed, 
corresponds to cluster-cluster aggregation class of models for $f_{b}=1$ 
\cite{jullien}, with a tendency towards cluster-cluster reaction limited 
aggregation process \cite{new_ref} for $f_{b}<1$.
This out of equilibrium phenomenon can be representative of colloidal 
gelation and will not be considered here. In chemical
gelation and colloidal gelation the formation of the critical cluster
should produce the slow dynamics.  The main difference is due to the fact that
the critical density and the temperature are much lower in colloidal gelation
than in chemical gelation, moreover the fractal dimension is related to
cluster-cluster aggregation models and not to random percolation.   

\section{Relaxation functions}  
\label{two}
In order to investigate the dynamic behavior we study, for both the 
permanent bond and the finite bond lifetime cases, the equilibrium 
density fluctuation autocorrelation functions, $f_{\vec{q}}(t)$, given by
\begin{equation}
f_{\vec{q}}(t) = \frac{< \rho_{\vec{q}} (t+t')
 \rho_{-\vec{q}}(t')>}{<|\rho_{\vec{q}}(t')|>^{2}}
\label{autct}
\end{equation}
where $\rho _{\vec{q}}(t)= \sum_{i=1}^{N} e^{-i \vec{q} \cdot \vec{r}_{i}(t)}$,
$\vec{r}_{i}(t)$ is
the position of the $i-th$ monomer at time $t$, $N$ is the number of monomers
and the average $\langle ... \rangle$ is performed over the time $t'$.
Due to periodic boundary conditions the values of the wave vector
$\vec{q}$ on the cubic lattice are $\vec{q}=\frac{2\pi}{L}(n_{x},n_{y},n_{z})$
with $n_{x},n_{y},n_{z}=1,~...,~ L/2$ integer values (in our simulations a cubic
lattice of size $L=16$ has been considered) \cite{nota_err1}.
In the following we discuss the data obtained for $q \sim 1.36$ ($\vec{q} = (
\pi/4, \pi/4, \pi/4$ )). Qualitatively analogous behaviors have been observed 
for ($\vec{q} = ( \pi/2, \pi/2, \pi/2$ ) and ($\pi, \pi, \pi$)).
Undoubtly, due to structure formation over different length scales, 
a detailed study of the geometric properties and the dynamics for different 
wave vectors might be relevant \cite{dedecon6}.
    
In the case of permanent bonds the system is considered at equilibrium when
both the diffusion coefficient has reached its asymptotic
limit, and the autocorrelation functions have gone to zero.
For $\phi<\phi_c$ we are always able to thermalize the system,
instead for $\phi>\phi_c$ it remains out of equilibrium, and it is possible
that it is in an aging regime \cite{dedecon6}. 
In Fig.\ref{fig2}, $f_{\vec{q}}(t)$  is plotted as function of the time for 
different values of the monomer concentration. The data 
have been averaged over $40$ different initial configurations.
The different curves correspond to different values of $\phi$, ranging from 
0.5 to 0.85. 
At low concentrations the system completely relaxes within the simulation time; 
the relaxation process becomes slower as the concentration is increased and 
above the percolation threshold, $\phi_{c}$, 
the system is kinetically arrested, in the sense that the relaxation functions 
do not go to zero within the time scale of the simulations. 

We analyze more quantitatively the long time decay of $f_{\vec{q}}(t)$ 
in Fig.\ref{fig3}: 
As the monomer concentration, $\phi$, approaches the percolation threshold, 
$\phi_{c}$, $f_{\vec{q}}(t)$ displays a long time decay well fitted by a 
stretched exponential law $\sim e^{-(t/\tau)^{\beta}}$ with a 
$\beta \sim 0.30 \pm 0.05$. Intuitively, this behavior can be related to the 
cluster size distribution close to the gelation threshold, which produces 
relaxation processes taking place over different length scales. 
At the percolation threshold the onset of a power 
law decay (with an exponent $c$) is observed as shown by the double logarithmic 
plot of Fig.\ref{fig3} \cite{relax}. 
As the monomer concentration is increased above the percolation threshold, i.e.
in the gel phase, the long time power law decay of the relaxation functions can 
be fitted with a decreasing exponent $c$, varying from $c\sim 1.0$ at 
$\phi_{c}$ to $c\sim0.2$ well above $\phi_{c}$, where a nearly logarithmic 
decay appears.  
This suggests that the growth of the relaxation time is driven by the
formation of the critical cluster, that actually determines the kinetic arrest.
On the whole, this behavior well agrees with the one observed in gelling 
systems investigated in the experiments of refs.\cite{relax}. 
It is interesting to 
notice that this kind of decay with a stretched exponential and a power law 
reminds the relaxation behavior found in spin-glasses \cite{ogi}.
Many analogies in the dynamics of gels and spin-glasses have been recently 
pointed out \cite{gelspin}, but the underlying physics is rather unclear. 

In the model with finite lifetime bonds, the equilibration time is an 
increasing function of $\tau_b$.  The system is considered at equilibrium when
both the average number of bonds has reached its asymptotic limit
and the autocorrelation functions have gone to zero \cite{nota_aging}.
In this case very different behaviors are 
observed. In Fig.\ref{fig4} 
$f_{\vec{q}}(t)$ is plotted as function of time 
for a fixed value of $\tau_{b}=10,100,1000$ (respectively Fig.\ref{fig4}$a$, 
\ref{fig4}$b$ and \ref{fig4}$c$) for increasing values of the 
monomer concentration ($\phi$ varies  on the same range as the permanent bond
case). 
At low concentrations, the behavior of the autocorrelation function 
$f_{\vec{q}}(t)$ 
is apparently very similar to the one observed in the case of permanent bonds: 
the system completely relaxes within the simulation time scale and the 
relaxation time increases with the concentration $\phi$. 
At high concentrations, a two-step decay appears, closely resembling the 
one observed in
super-cooled liquids. This qualitative behavior is observed for many 
different values of the bond lifetime, $\tau_{b}$. As shown in Fig.
\ref{fig4}, the two step decay is more pronounced for higher 
values of $\tau_{b}$. 

As we can see in Fig.\ref{fig5}, where the long time decay of $f_{\vec{q}}(t)$
for $\tau_{b}=100$ is shown, the long time decays are well fitted by 
stretched exponentials.
The exponent $\beta$ ($\beta \sim 0.7$ for the case of Fig.\ref{fig5}) does 
not seem to vary significantly as the concentration
varies, and this has been observed for all the values of $\tau_{b}$ studied. 
It instead decreases as $\tau_{b}$ increases: for very small $\tau_{b}$ one 
recovers a long time exponential decay whose behavior
becomes less and less exponential as the bond lifetime increases. This suggests 
that the stretched exponential decay is due to the presence of 
long living structures. 

For high monomer concentrations we fit the $f_{\vec{q}}(t)$ curves  
using the mode-coupling $\beta$-correlator \cite{goe}, 
corresponding to a short time power law 
$\sim f + \left( \frac{t}{\tau_{s}} \right)^{-a}$ and a long time von 
Schweidler law $\sim f - \left( \frac{t}{\tau_{l}} \right) ^{b}$. 
In Fig.\ref{fig6} we show the agreement between the fit (the full lines) 
and the data for $\tau_{b}=1000$ in the range of concentrations $\phi=0.8-0.9$. 
The exponents obtained are $a \sim 0.33 {\pm} 0.01$ and 
$b \sim 0.65 {\pm} 0.01$. At long times the different curves obtained for 
different $\phi$ collapse onto a unique master curve by opportunely rescaling 
the time via a factor $\tau(\phi)$ (Fig.\ref{fig7}). 
The master curve is well fitted by a stretched exponential decay with 
$\beta \sim 0.50 {\pm} 0.06$. The characteristic time $\tau(\phi)$ diverges at 
a value $\phi_{g} \sim 0.963 {\pm}0.005$ with the exponent 
$\gamma \sim 2.33 {\pm} 0.06$ (Fig.\ref{fig8}). 
This value well agrees with the mode-coupling 
prediction $\gamma = 1/2a + 1/2b$ \cite{goe}.

The same behavior and the same level of agreement between the data and 
the mode coupling predictions have been obtained for different values of 
$\tau_{b}$ ($\tau_{b}=100, 200, 400, 1000, 3000$) . 
Neither the exponents $a$ and $b$ obtained by the 
$\beta$-correlator nor the exponent $\beta$ of the stretched exponential 
vary significantly as function of $\phi$ and of $\tau_{b}$. The value of 
$\phi_{g}$, where the characteristic time $\tau(\phi)$ apparently diverges, 
seems instead to vary  with $\tau_{b}$, but it is always close to
$\phi_{g}=1$. 

This glassy relaxation pattern suggests that also in this case the relaxation 
takes place by means of a caging mechanism: particles are trapped in a cage 
formed by the surrounding ones, the first relaxation step is due to movements 
within this cage, whereas the final relaxation is possible due to 
cage opening and rearrangement.
We notice that, contrary to the usual behavior observed in super-cooled 
liquids and predicted by the Mode-Coupling Theory, the value of the plateau of 
the relaxation functions, which is typically related to the size of the cage, 
strongly depends on the monomer concentration, $\phi$. This effect will be 
explained later in terms of effective clusters.

\section{The relaxation time and the role of effective clusters }
\label{three}
We study now the relaxation times that can be obtained from 
the $f_{\vec{q}}(t)$, as the time $\tau$ such that $f_{\vec{q}}(\tau) \sim
0.1$.  In Fig.\ref{fig9} the relaxation time $\tau$ 
is plotted as function of the 
monomer concentration, $\phi$, for the permanent bonds and for the
finite lifetime bonds case at different values of $\tau_b$. In the figure 
one finds the data for the permanent bond case on the left, and then from left 
to right the data for decreasing values of bond lifetime, $\tau_{b}$.

In the case of permanent bonds (chemical gelation), 
$\tau(\phi)$ displays a power law divergence 
at the percolation threshold $\phi_{c}$. For finite bond lifetime 
the relaxation time instead increases following the permanent bond case, 
up to some value $\phi^{*}$ and then deviates from it.  
The longer the bond lifetime the higher $\phi^{*}$ is. For higher $\phi$ 
the further increase of the relaxation time corresponds to the onset of the 
glassy regime in the relaxation behavior discussed in the previous section.  
This truncated critical behavior followed by a glassy-like transition has been 
actually detected in some colloidal systems in the viscosity behavior 
\cite{malla,durand}. 

In both cases of permanent bonds and  finite lifetime bonds, 
clusters of different sizes are present in the system. 
In the permanent bond case, a cluster of radius $R$ diffuses in the medium 
formed by the other percolation clusters with a characteristic relaxation 
time $\tau(R)$. At the percolation threshold the 
connectedness length critically grows in the system and so does the overall 
relaxation time. 
In the case of a finite bond lifetime $\tau_{b}$, it will exist a cluster size
$R^{*}$ so that $\tau_{b} < \tau(R^{*})$. That is, clusters of size
$R \geq R^{*}$ will break and lose their identity on a time scale shorter
than $\tau(R)$ and their full size will not contribute to the enhancement 
of the relaxation time in the system. We can say that 
the finite bond lifetime actually introduces an effective cluster size 
distribution with a cut-off and keeps the macroscopic viscosity finite in the 
system \cite{conmess}.

At high concentrations the system approaches a glassy regime and the 
relaxation time increases.
In order to further investigate the high concentration regime, in 
Fig.\ref{fig10} 
we directly compare $f_{\vec{q}}(t)$ at fixed $\phi=0.85$
for $\tau_{b}=10,100,400,1000$, and the permanent bond case. 
We observe that at a fixed value of the monomer 
concentration, $\phi$, the relaxation functions calculated 
for finite lifetime bonds coincide with the permanent bond case 
up to times of the order of $\tau_{b}$. 
This suggests that on time scales smaller 
than $\tau_{b}$ the relaxation process must be on the whole the same as in the 
case of permanent bonds, where permanent clusters are present in the system,
and gives an interpretation in terms of effective clusters for the two
step glassy behavior of the relaxation functions:                               
The first step is due to the relaxation of a cluster
within the cage formed by the other clusters, whereas the second step is due
to the breaking of clusters. This second relaxation is the analog of the cage
opening in an ordinary supercooled liquid. In conclusion, on a time scale of
the order of $\tau_b$, the effective clusters play the role of single molecules
in an ordinary supercooled liquid, or in a colloidal hard sphere system.        

Using this picture of effective clusters, we are able to explain the
increase of the plateau in $f_{q}(t)$. In fact, since different values of the 
monomer concentration correspond to  different 
effective cluster size distributions, for each value of $\phi$ one has a 
different glassy liquid of effective clusters. 
This will change the first relaxation and should 
correspond to a change in the plateau of the relaxation functions 
(Fig.\ref{fig4} and \ref{fig6}). In particular we find that for higher 
$\phi$ one has a higher plateau, that is the first decay 
(the motion of clusters within the cages) produces a smaller relaxation in 
the system.

\section{Diffusion properties}
\label{four}
In order to obtain further information on the dynamics 
we calculate the mean square displacement of all the particles,
$\langle \vec{r}^{2}(t) \rangle = \frac{1}{N} \sum_{i=1}^{N} \langle
(\vec{r}_{i}(t+t') - \vec{r}_{i}(t') )^{2} \rangle$. In the model with
finite lifetime bonds, clusters continuously evolve in time
and therefore the diffusion coefficient of a single cluster cannot be defined.

In the model with permanent bonds the mean square displacement of the particles
$\langle \vec{r}^{2}(t) \rangle$ 
presents a long-time diffusive behavior, and the diffusion coefficient 
decreases but remains finite also above $\phi_{c}$. This is due to the fact
that the infinite cluster can be viewed as a net with a large mesh size,
through which monomers and small clusters can diffuse.

In previous papers \cite{dedecon3} the diffusion coefficient of clusters with
a fixed size was studied. We found that  clusters, whose size is comparable 
with the connectedness length, present a diffusion coefficient going to zero 
at $\phi_{c}$ (with the same exponent as the relaxation time), whereas single 
monomers present a finite diffusion coefficient also in the gel phase. 
As we have already noticed, this is due the fact that small 
clusters are able to escape through the percolating cluster having a  
structure with holes over many different length scales close to the percolation 
threshold. It is therefore clear that the behavior here observed for the mean 
square displacement is mainly due to the diffusion of single monomers and small
clusters. 
In Fig.\ref{fig11}, $\langle \vec{r}^{2}(t) \rangle$ is plotted as function of 
time in a double logarithmic plot for $\phi$ approaching $\phi_{c}$ 
in the case of permanent bonds: 
particles can still diffuse, and the diffusion coefficient apparently decreases 
with the monomer concentration. At high concentrations 
the sub-diffusive regime stays longer, and the long time diffusive behavior 
is hardly recovered. 

In Fig.\ref{fig12}, we plot the data 
obtained with $\tau_{b}=1000$ and $\phi=0.8, 0.82, 0.85, 0.9$.  
According to the results just discussed, for low concentrations 
$\langle \vec{r}^{2}(t)\rangle$ shows a simple diffusive behavior.
The diffusion does not change significantly close to $\phi_{c}$ and for high 
concentrations the behavior observed, characterized by a plateau, is similarly 
to glass forming systems. This onset of a glassy regime has been 
obtained for different values of $\tau_{b}$, and again it indicates a caging 
mechanism in the dynamics. The asymptotic diffusion coefficient goes to zero 
as $\phi$ approaches $\phi_{g}$ (inset of Fig.\ref{fig12}), as a power law, 
with an exponent close to $\gamma$ (section \ref{three}) 
in agreement with the Mode Coupling Theory predictions. 

As already done for the relaxation functions we directly compare the mean 
square displacement obtained in the cases of permanent bonds and finite bond 
lifetime. Fig.\ref{fig13} shows for a fixed value of the concentration,
$\phi=0.85$, that the two quantities 
coincide up to time scales of the order of $\tau_{b}$. 
For longer times in the system with non permanent bonds the 
final diffusive regime is recovered. These results are coherent with 
the behavior of the relaxation functions discussed in the previous section.
The first regime is apparently related to the diffusion of effective clusters. 
Here again the value of the plateau in the diffusion pattern, which is related
to the size of the cage, varies with the concentration $\phi$. For higher 
values of 
the monomer concentration, the size of the cage apparently decreases. 
This corresponds to larger effective clusters, which have less free space  
compared to their size. By means of the qualitative argument used in 
section \ref{three}, one expects that for a longer $\tau_{b}$ the condition 
$\tau_{b} < \tau(R^*)$ will be fulfilled by a larger size $R^*$, and on  
average larger clusters will persist.   
For the same value of the concentration, the size of 
the cage should be the same, whereas the {\em particles} of this glassy system 
(i.e. the effective clusters) are longer trapped in the cage as the 
bond lifetime increases (Fig.\ref{fig13}).   

\section{Discussion and Conclusion}
We have studied a model for gelling systems both in the case of permanent 
bonds and finite bond lifetime. The study of the dynamics shows that 
when bonds are permanent (chemical gelation) the divergence of the relaxation 
time is due to the formation of a macroscopic critical cluster and the decay 
of the relaxation functions is related to the relaxation of such  cluster. 
In the case of finite $\tau_{b}$ there is an effective cluster size 
distribution, with a size cutoff.
Note  that the clusters cannot be easily defined as in the case of chemical
gelation: The effective clusters do not coincide with pairwise bonded particles.
A cluster can be identified in a statistical sense as a group of monomers which
keeps its identity (i.e. the bonds are unbroken) when diffusing a distance of 
the order of its diameter. 
The  formation of effective clusters leads to an apparent divergence of the
relaxation time which is characterized by exponents corresponding to the
case of random permanent bonds (random percolation).
As the monomer density increases the presence of effective clusters further
slows down the dynamics, until a glass transition is reached. 

In the case $\tau_b\rightarrow \infty$, starting with a random
configuration of unbonded monomers one obtains an out of equilibrium state
as in cluster-cluster aggregation models,  
which can be representative of colloidal gelation.  
Ideally this out of equilibrium system is connected to the two lines described 
above, the pseudo percolation line
and the glassy line. The pseudo percolation line can be detected 
if the effective
cluster size is large enough and it is quite distinct from the glassy line.
However both lines interfere at low densities and low temperatures with the 
phase coexistence curve.

We would like to thank 
K. Dawson, A. de Candia, G. Foffi, W. Kob, F. Mallamace, N. Sator, 
F. Sciortino, P. Tartaglia and E. Zaccarelli for many interesting discussions.
This work has been partially supported by a Marie Curie Fellowship of the 
European Community programme FP5 under contract number HPMF-CI2002-01945,       by MIUR-PRIN 2002, MIUR-FIRB 2002, CRdC-AMRA, and by the INFM Parallel 
Computing Initiative.
%
%
 
%
\newpage
\begin{figure}
\begin{center}
\mbox{ \epsfxsize=8cm
       \epsfysize=7cm
       \epsffile{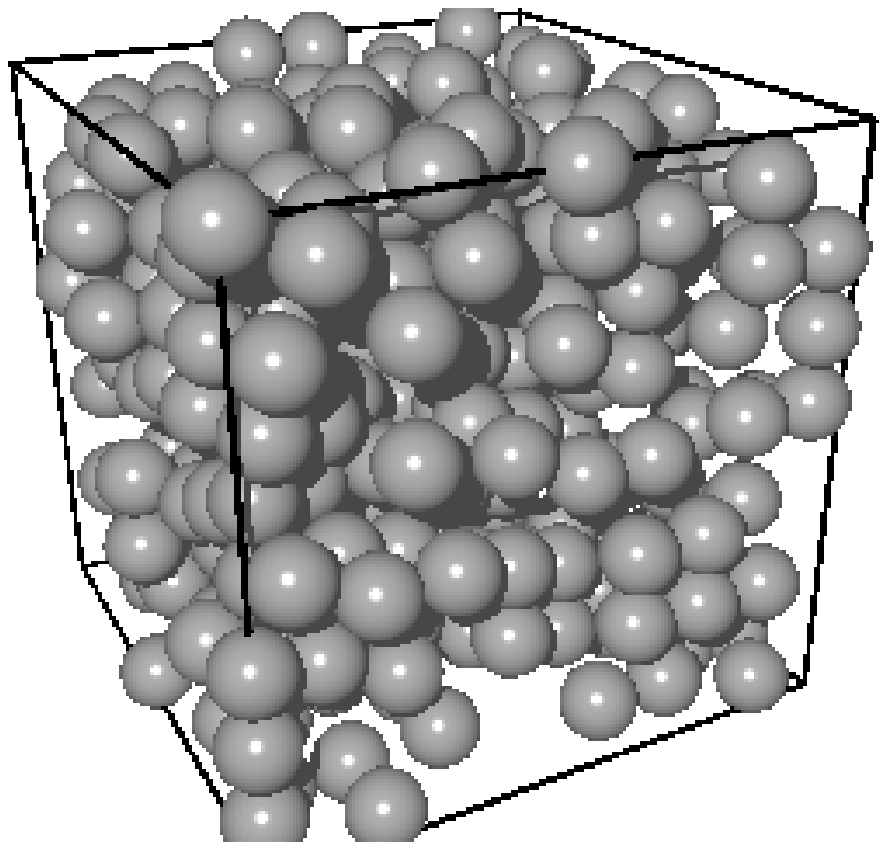}
       }
\end{center}
\end{figure}

\begin{figure}
\begin{center}
\mbox{ \epsfxsize=8cm
       \epsfysize=7cm
              \epsffile{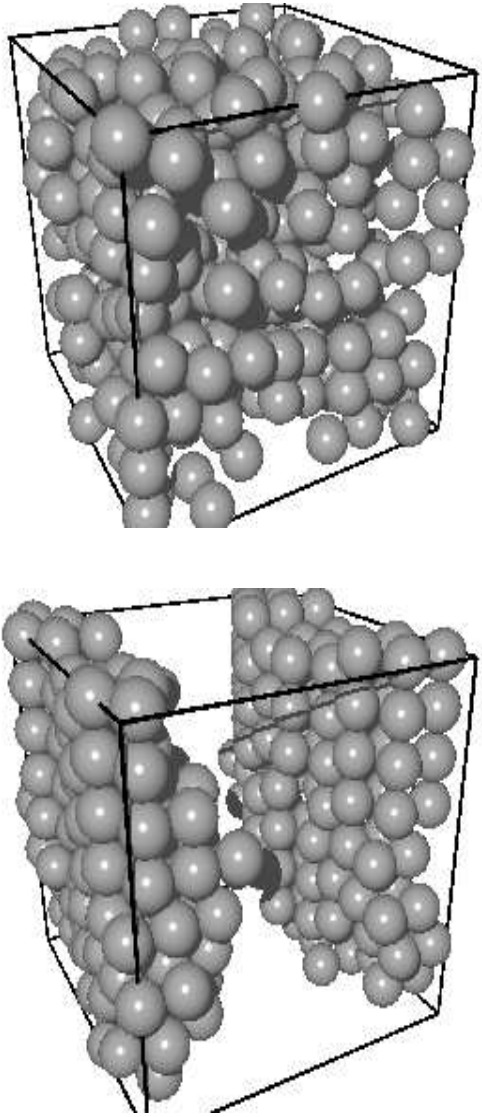}
	             }
\end{center}
\caption{
Two typical configurations obtained for $\phi=0.6$
with  $\tau_{b}=100$ and $f_{b}=0.02$ ($a$), 
where there is no evidence of phase separation, and with $\tau_{b}=2$,  
$f_{b}=1$ and monomers of valence 6 ($b$), where the phase 
separation occurs. 
}
\label{fig1}
\end{figure} 

\begin{figure}
\begin{center}
\mbox{ \epsfxsize=8cm
       \epsfysize=8cm
       \epsffile{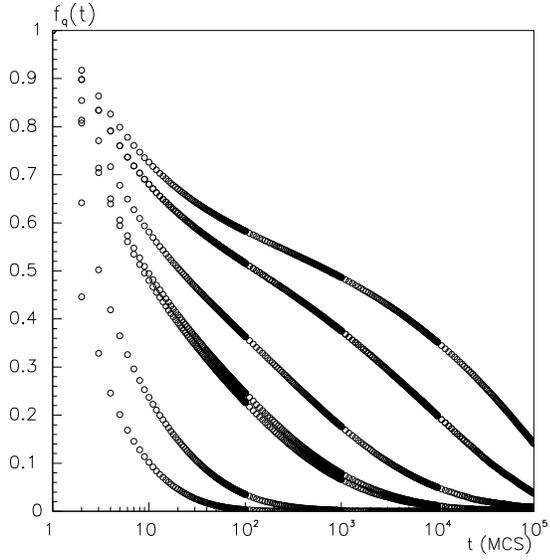}
      }
\end{center}
\caption{
The relaxation functions for permanent bonds 
$f_{\vec{q}}(t)$ as function of the time for $q \sim 1.36$ and, from left to 
right, $\phi= 0.5, 0.6, 0.68, 0.718, 0.75, 0.8, 0.85$. 
}
\label{fig2}
\end{figure}               

\begin{figure}
\begin{center}
\mbox{ \epsfxsize=8cm
       \epsfysize=8cm
       \epsffile{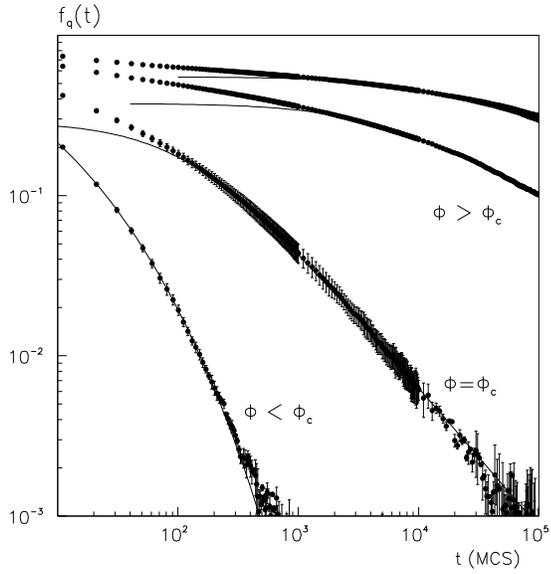}
      }
\end{center}
\caption{
Double logarithmic plot of the autocorrelation functions
$f_{\vec{q}}(t)$ as function of the time for $q \sim 1.36$ and
$\phi=0.6, 0.718, 0.8, 0.87$.
For $\phi < \phi_{c}$ the long time decay is well fitted by a
function (full line) $\sim e^{-(t/\tau)^{\beta}}$ with $\beta \sim 0.3$.
At the percolation threshold and in the gel phase in the long time decay the
data are well fitted by a function $\sim (1 + \frac{t}{\tau'})^{-c}$.
}      
\label{fig3}
\end{figure}                

\begin{figure}
\begin{center}
\mbox{ \epsfxsize=8cm
       \epsfysize=8cm
       \epsffile{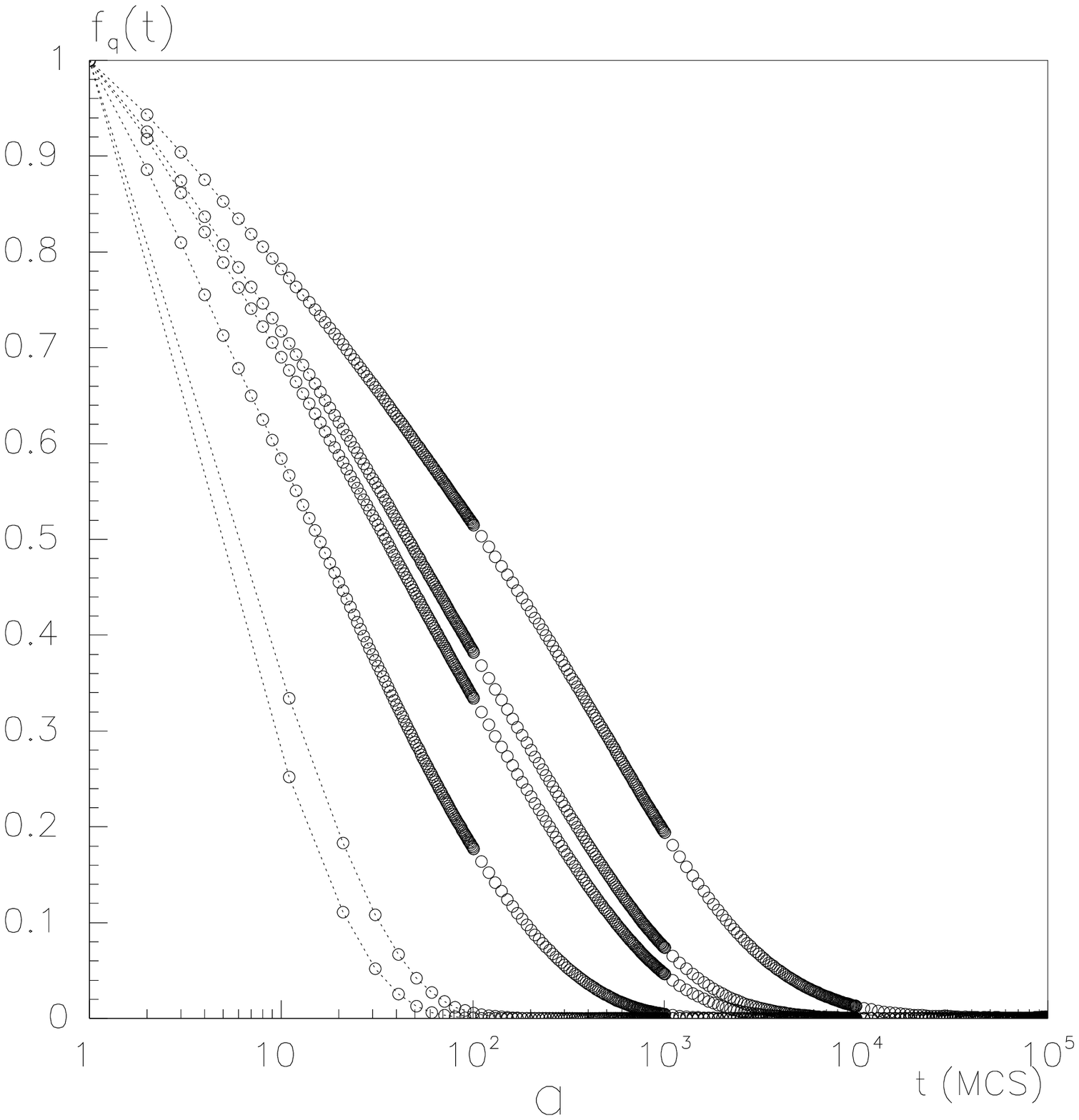}
                 }
\end{center}
\end{figure}
					  
\begin{figure}
\begin{center}
\mbox{ \epsfxsize=8cm
       \epsfysize=8cm
       \epsffile{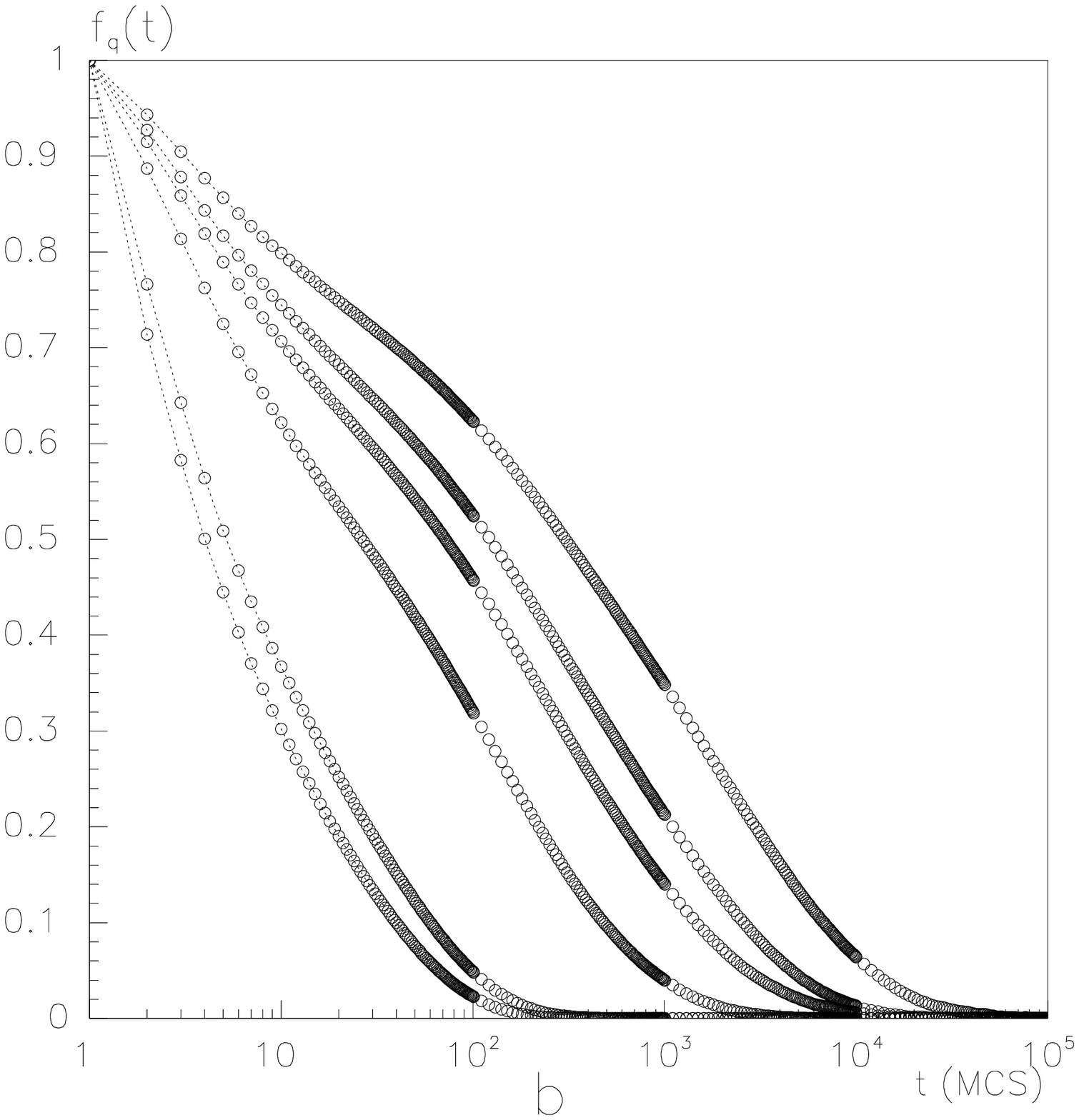}
                }
\end{center}
\end{figure}
					  
\begin{figure}
\begin{center}
\mbox{ \epsfxsize=8cm
       \epsfysize=8cm
       \epsffile{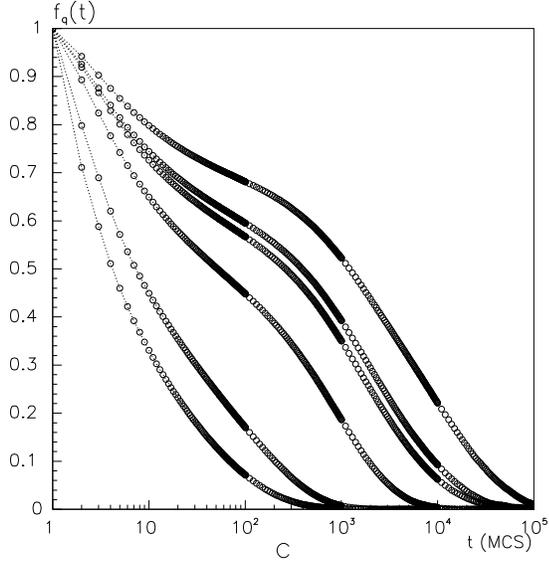}
      }
\end{center}
\caption{
$f_{\vec{q}}(t)$ as function of the time for $q \sim 1.36$
calculated on a cubic lattice of size $L=16$: for
$\phi=0.6,0.7,0.8,0.85,0.87,0.9$ (from left to right) and
$\tau_{b}=10 MCstep/particle$ 
($a$); 
$\tau_{b}=100 MCstep/particle$
($b$);
$\tau_{b}=1000 MCstep/particle$ 
($c$);
the dotted lines are a guide to the eye.
}
\label{fig4}
\end{figure}    

\begin{figure}
\begin{center}
\mbox{ \epsfxsize=8cm
       \epsfysize=8cm
       \epsffile{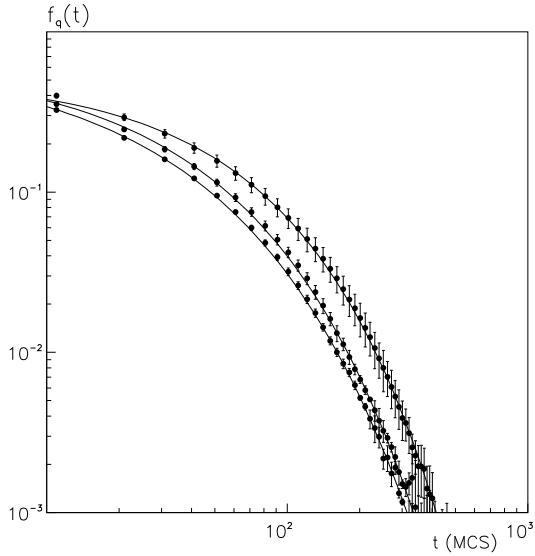}
      }
\end{center}
\caption{
The long time decay of 
$f_{\vec{q}}(t)$ in a log-log plot for $q \sim 1.36$. 
It has been calculated on a cubic lattice of size $L=16$ for
$\tau_{b}=100 MCstep/particle$ (from left to right $\phi=0.65,0.68,0.718$). 
The data are fitted using a stretched exponential function with $\beta \sim 0.7$
(full lines).}
\label{fig5}
\end{figure} 

\begin{figure}
\begin{center}
\mbox{ \epsfxsize=8cm
       \epsfysize=8cm
       \epsffile{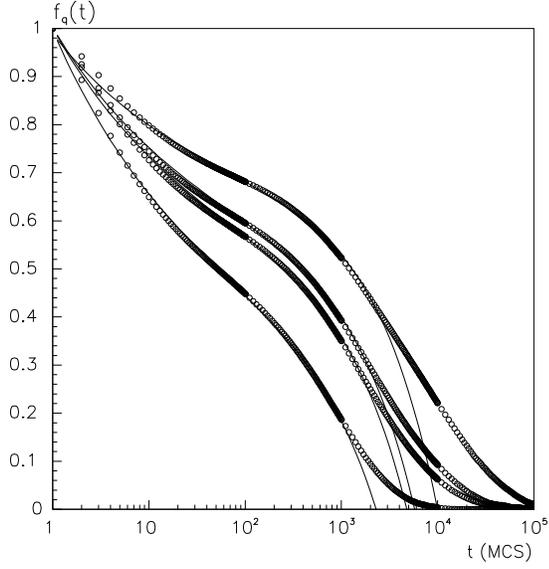}
      }
\end{center}
\caption{
$f_{\vec{q}}(t)$ as function of the time for $q \sim 1.36$
calculated on a cubic lattice of size $L=16$ for
$\tau_{b}=1000 MCstep/particle$. From left to right 
$\phi= 0.8, 0.85, 0.87, 0.9$. The full lines correspond to the fit 
with the $\beta$-correlator. 
}
\label{fig6}
\end{figure} 

\begin{figure}
\begin{center}
\mbox{ \epsfxsize=8cm
       \epsfysize=8cm
       \epsffile{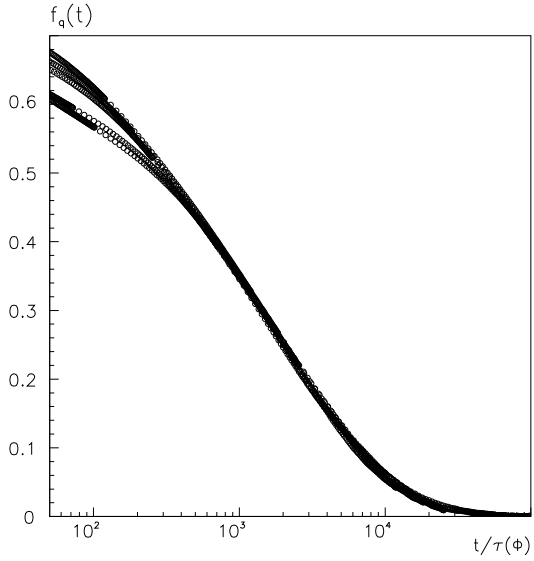}
      }
\end{center}
\caption{
$f_{\vec{q}}(t)$ obtained for $q \sim 1.36$, $\tau_{b}=1000$, and 
$\phi= 0.85, 0.87, 0.9, 0.91, 0.92$: by opportunely rescaling them by a 
quantity $\tau(\phi)$ they collapse into a unique master curve, well fitted by
a stretched exponential function with $\beta \sim 0.5 \pm 0.06$. 
}
\label{fig7}
\end{figure} 

\begin{figure}
\begin{center}
\mbox{ \epsfxsize=8cm
       \epsfysize=8cm
       \epsffile{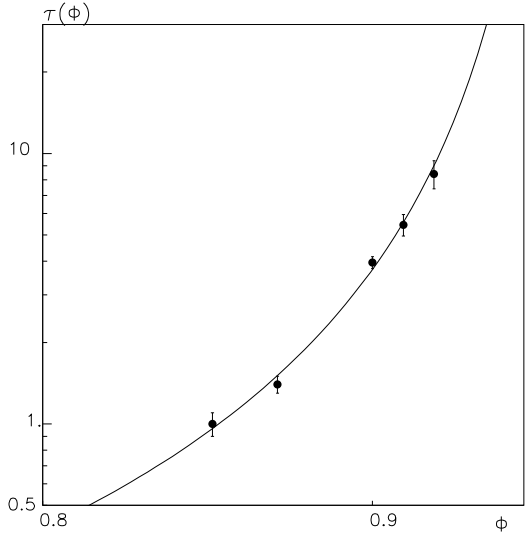}
      }
\end{center}
\caption{
Log-log plot of the characteristic time $\tau(\phi)$ obtained by the 
rescaling of the relaxation functions. The points have been fitted (full line) 
with the function $0.006(\phi_{g} - \phi)^{-2.33}$, where $\phi_{g} \sim 0.96 
\pm 0.01$ .}
\label{fig8}
\end{figure} 

\begin{figure}
\begin{center}
\mbox{ \epsfxsize=8cm
       \epsfysize=8cm
       \epsffile{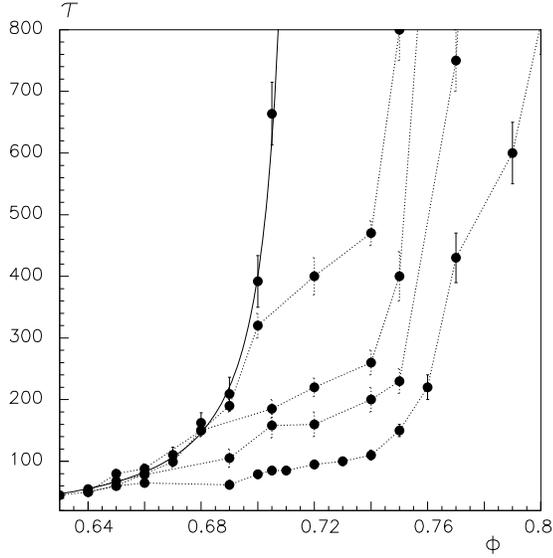}
      }
\end{center}
\caption{
The average relaxation time as function of the density; from left to right:
the data for the permanent bond case
diverge at the percolation threshold with a power law (the full line);
the other data refer to finite $\tau_{b}=3000,1000,400,100 MC step/particle$
decreasing from left to right (the dotted lines are a guide to the eye).
The apparent divergence of the relaxation time, $\tau$, is observed at the
percolation threshold of the permanent bond case, $\phi_c=0.718$, for all the
values of $\tau_b$.
}    
\label{fig9}
\end{figure}

\begin{figure}
\begin{center}
\mbox{ \epsfxsize=8cm
       \epsfysize=8cm
       \epsffile{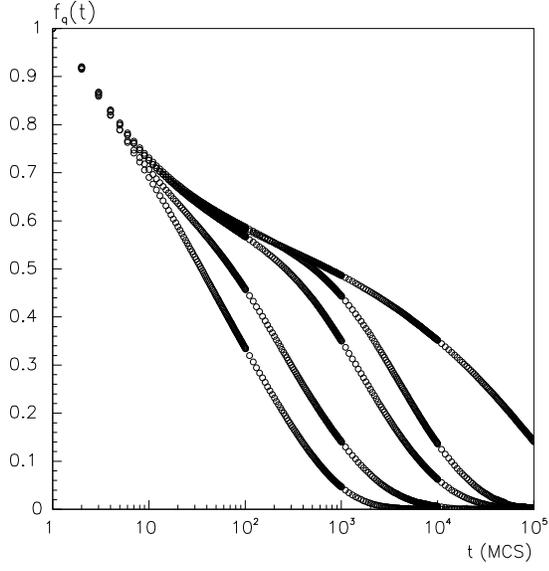}
      }
\end{center}
\caption{
$f_{\vec{q}}(t)$ obtained for $\phi=0.85$ and $q \sim 1.36$: the different 
curves refer to $\tau_{b}=10, 100, 400, 1000$, compared to the permanent bond
case (from left to right).
}
\label{fig10}
\end{figure}  

\begin{figure}
\begin{center}
\mbox{ \epsfxsize=8cm
       \epsfysize=8cm
       \epsffile{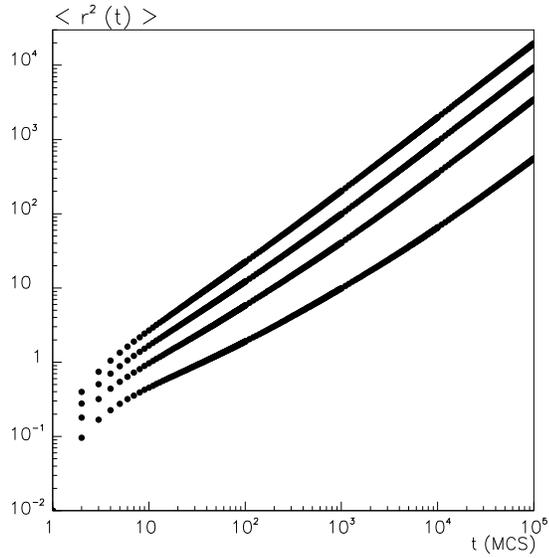}
      }
\end{center}
\caption{
The mean-square displacement $\langle \vec{r}^{2}(t) \rangle$ of the
particles as function of the time in a double logarithmic plot for
permanent bonds: from top to
bottom $\phi=0.4,0.5,0.6,0.7$, approaching $\phi_{c}$.
} 

\label{fig11}
\end{figure}  

\begin{figure}
\begin{center}
\mbox{ \epsfxsize=8cm
       \epsfysize=8cm
       \epsffile{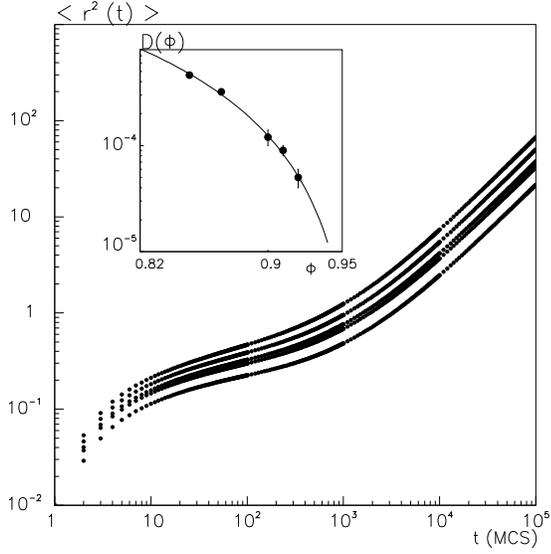}
      }
\end{center}
\caption{
The mean-square displacement $\langle \vec{r}^{2}(t) \rangle$ of the
particles as function of the time in a double logarithmic plot for
$\tau_{b} = 1000 MCstep/particle$: from top to
bottom $\phi=0.8,0.82,0.85,0.87,0.9$, approaching $\phi_{g}(\tau_{b})$.
In the inset, the diffusion coefficient: the full line is the fit with the 
function $\sim (0.963-\phi)^{-2.3}$. }  
\label{fig12}
\end{figure}   

\begin{figure}
\begin{center}
\mbox{ \epsfxsize=8cm
       \epsfysize=8cm
       \epsffile{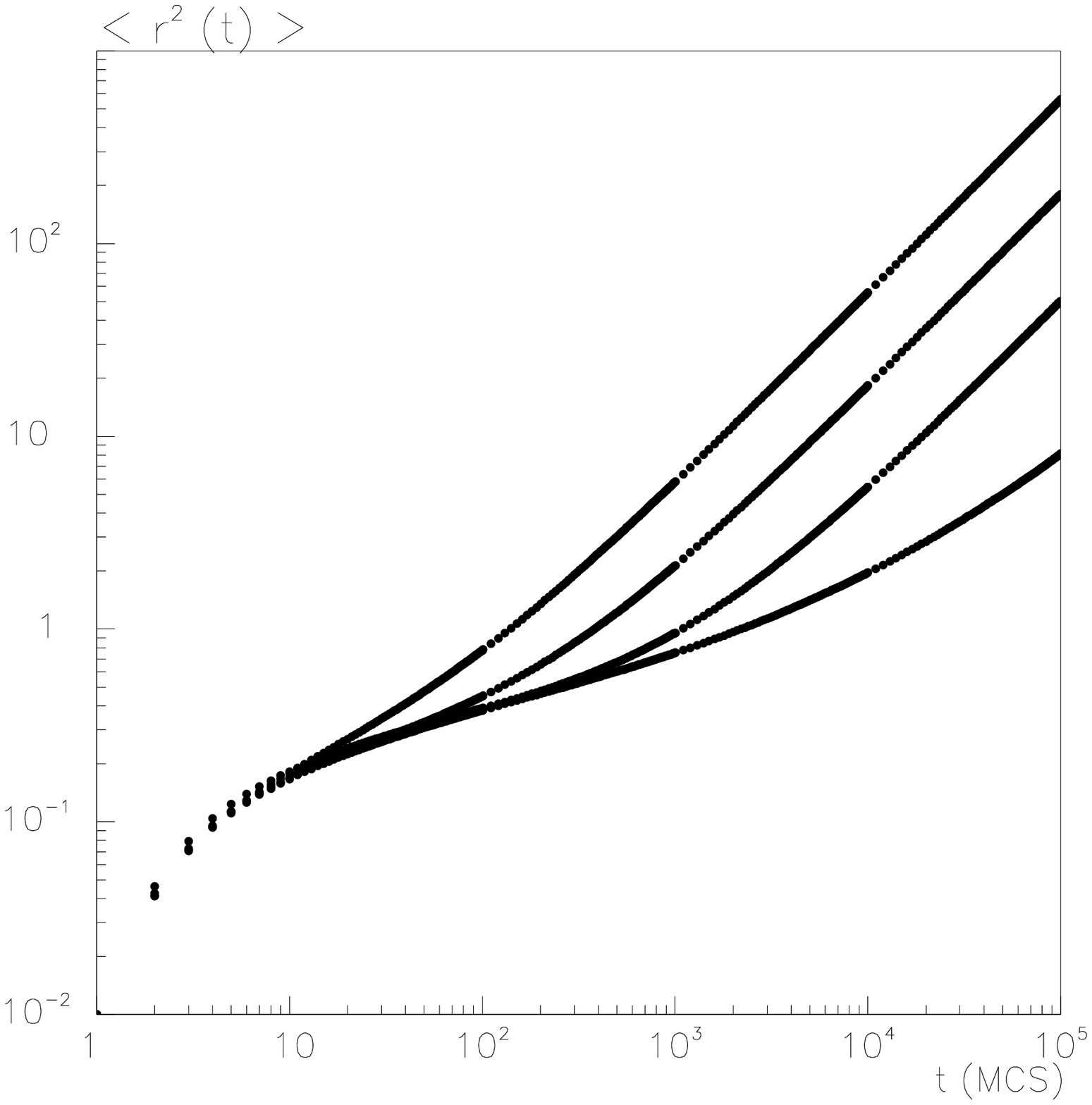}
      }
\end{center}
\caption{
The mean-square displacement $\langle \vec{r}^{2}(t) \rangle$ of the
particles as function of the time in a double logarithmic plot, obtained at 
$\phi=0.85$. The different curves, from top to bottom, refer to 
$\tau_{b}=10, 100, 1000$ and the case of permanent bonds.
}
\label{fig13}
\end{figure}


\begin{thebibliography}{999}
\bibitem{nota3}
Taking into account both the interaction energy by means of $\tau_b$ and the
frequency $f_b$ for the bonds corresponds to have a free energy barrier in the
bond formation \cite{constkl}: 
Consider a pair of particles which can be in $\Omega + 1$ configurations.
$\Omega$ (unbonded) corresponds to zero energy and $1$ (bonded)  corresponds to
energy $- E$. Breaking of a bond corresponds to go from a state of energy $-
E$ to a state of entropy $S=K \ln \Omega$, with a variation in the free energy
$\Delta F=E-KT\ln \Omega$. The 
lifetime of the bonding configuration is proportional to $e^{E/KT}$, and 
the frequency to go from one of the unbonding configurations
to the bonding configuration is given by $f_b=\frac{1}{\Omega}$.
\bibitem{nota1}
We consider one starting configuration with a fixed $\phi$ as obtained in the
case of permanent bonds. For each value of the bond lifetime $\tau_{b}$ we
perform test simulations for different values of $f_{b}$. The fraction
of formed bonds respect to all the possible ones varies during an initial
transient and then fluctuates around an average value which, 
for a fixed $\phi$, depends on $\tau_{b}$ and $f_{b}$. 
For the study of the dynamics, for each value
of $\tau_{b}$ we take the $f_{b}$ so that the average value of the fraction of
present bonds during the simulations is the same as its value 
at the same $\phi$ with permanent bonds.
\bibitem{nota2}
The role of directional effects in the attractive interaction in colloidal gels
is not clear and is worth to be further investigated.
\bibitem{nota_err1}  
After having thermalized the system at a given value of the monomer
concentration, $\phi$,
the autocorrelation functions of the density fluctuations, $f_{\vec{q}}(t)$, 
are calculated by time averages over runs of $\sim 10^7 MCstep/particle$.
The errors are calculated as the
fluctuations on the statistical ensemble ($\sim 20 - 30 $ different realizations
of the system have been considered). 
Where not explicitly shown, the error bars are of the order of the
symbol size in the figures.
\bibitem{nota_aging}
Although with these two conditions we cannot exclude that we are in an aging 
regime, we do not expect to find aging, as we are not in the glassy 
region \cite{dedecon6}.
\end{thebibliography}
\end{document}